\documentclass[aps,twocolumn,prx,amsfonts,amssymb,showpacs,superscriptaddress,floatfix, longbibliography]{revtex4-1}

\usepackage[latin1]{inputenc}
\usepackage{graphicx}
\usepackage{dcolumn}
\usepackage{bm}
\usepackage{subfigure}
\usepackage{float}
\usepackage{multirow}
\usepackage{amsmath}
\usepackage{graphicx}
\usepackage{color}
\usepackage{hyperref}
\usepackage{psfrag}
\usepackage{amssymb}
\usepackage{wasysym}
\usepackage{mathrsfs}
\usepackage[english]{babel}

\usepackage{times}

\begin{document}

\setlength{\baselineskip}{0.4cm} \addtolength{\topmargin}{1.5cm}

\title{Continuously Sheared Granular Matter Reproduces in Detail Seismicity Laws}

\author{S. Lherminier}
 \affiliation{Institut Lumi\`ere Mati\`ere, UMR5306 Universit\'e Lyon 1-CNRS, Universit\'e de Lyon 69622 Villeurbanne, France.}
 
\author{R. Planet}
 \altaffiliation{Present address: Departament de F\'isica de la Mat\`eria Condensada, Universitat de Barcelona, Mart\'i i Franqu\`es 1, E-08028 Barcelona, Spain and Universitat de Barcelona, Institute of Complex Systems, Mart\'i i Franqu\`es 1, E-08028 Barcelona, Spain.}
 \affiliation{Institut Lumi\`ere Mati\`ere, UMR5306 Universit\'e Lyon 1-CNRS, Universit\'e de Lyon 69622 Villeurbanne, France.}
 
 \author{V. Levy dit Vehel}
 \affiliation{Institut Lumi\`ere Mati\`ere, UMR5306 Universit\'e Lyon 1-CNRS, Universit\'e de Lyon 69622 Villeurbanne, France.}
  
\author{G. Simon}
 \affiliation{Institut Lumi\`ere Mati\`ere, UMR5306 Universit\'e Lyon 1-CNRS, Universit\'e de Lyon 69622 Villeurbanne, France.}
 
\author{L. Vanel}
 \affiliation{Institut Lumi\`ere Mati\`ere, UMR5306 Universit\'e Lyon 1-CNRS, Universit\'e de Lyon 69622 Villeurbanne, France.}
 
\author{K. J. M\aa{}l\o{}y}
 \affiliation{PoreLab,  Department of Physics, University of Oslo, P. O. Box 1048, 0316 Oslo, Norway.}
 
\author{O. Ramos}
\email{osvanny.ramos@univ-lyon1.fr}
 \affiliation{Institut Lumi\`ere Mati\`ere, UMR5306 Universit\'e Lyon 1-CNRS, Universit\'e de Lyon 69622 Villeurbanne, France.}

\date{\today}

\begin{abstract}

We introduce a shear experiment that  \textit{quantitatively} reproduces the main laws of seismicity. By continuously and slowly shearing a compressed monolayer of disks in a ring-like geometry, our system delivers events of frictional failures with energies following a Gutenberg-Richter law. Moreover foreshocks and aftershocks are described by Omori laws and inter-event times also follow exactly the same distribution as real earthquakes, showing the existence of memory of past events. Other features of real earthquakes qualitatively reproduced in our system are both the existence of a quiescence preceding mainshocks, as well as magnitude correlations linked to large quakes. The key ingredient of the dynamics is the nature of the force network, governing the distribution of frictional thresholds. 

\end{abstract}

\maketitle

For more than a century, fracture and stick-slip frictional sliding have tried to explain the behavior of earthquakes. Brittle fracture induced by shear \cite{Reid1911} was the most accepted model until the sixties. However, a more precise analysis of the radiated waves \cite{Benioff1964}, the low amount of stress released by an earthquake in relation to the available one, the high energies needed to shear over a fractured surface, and over all, the lack of healing required to generate a second earthquake at the same location and close in time to the first one, set stick-slip sliding mechanisms as a more plausible explanation of earthquakes \cite{Brace1966}. 
Despite these facts, the subcritical fracture of heterogeneous materials shows naturally a jerky behavior that seems  closer to earthquake statistics than frictional sliding, which commonly displays a quasi-periodic stick-slip dynamics.
Indeed, several fracture experiments \cite{Scholz1968, Maloy2006, Baro2013, Stojanova2014, Bares2018} and numerical models \cite{Kun2013, Kun2014} have reported statistics of events following power-law distributions of sizes that have been compared to the the Gutenberg-Richter law \cite{Gutenberg1956}. The existence of aftershocks that follow the Omori law \cite{Omori1894} are also common in fracture experiments \cite{Maloy2006, Baro2013, Stojanova2014, Bares2018}.

\begin{figure}[b!]
\centering
\vspace{-0.5 cm}
\includegraphics[width=1.0\linewidth]{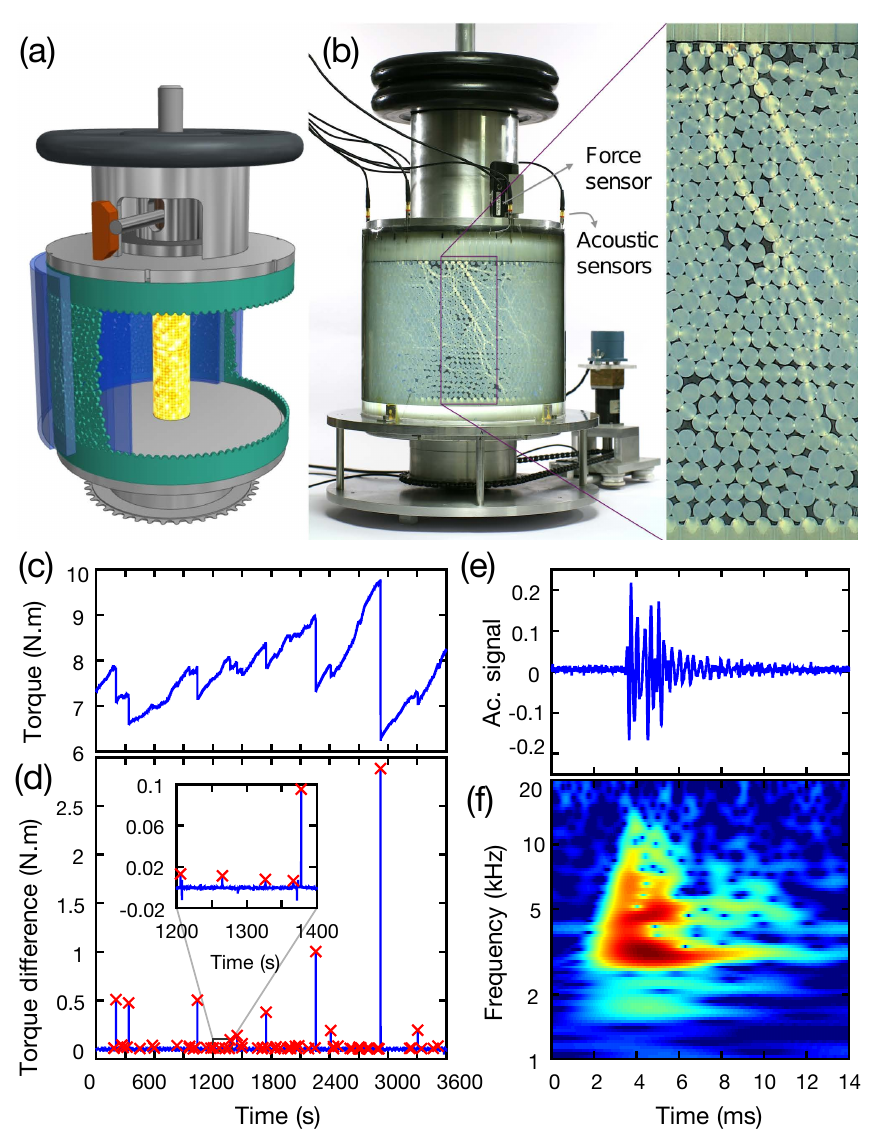}
\vspace{-0.5 cm}
	\caption{(color online) Setup and raw data analysis.
	(a) Sketch of the setup.
	(b) Photograph of the setup, displaying the mechanical and acoustic sensors. Inset: force chains in the granular layer observed thanks to photoelasticity.
	(c, d) Respectively torque signal and torque difference on 0.1 s intervals on a 1 hour window.
	Detected torque drops have been highlighted by {$\times$}.
	(e) Typical acoustic event.
	(f) Result of the Discrete Wavelet Transform (DWT) on the acoustic event, resulting in a time-frequency energy distribution map, with a color proportional to the logarithm of the energy value. }
	\label{fig:setup}
\end{figure}

Concerning stick-slip frictional sliding, different laboratory experiments have analyzed the sliding dynamics between two solid blocks. From a physical perspective, studies on acrylic blocks have focused on the complex evolution of the frictional strength during the slipping process, describing the behavior as a dynamic fracture problem \cite{Rubinstein2004, Ben-David2010}. Recent friction experiments on rocks have reported results on supershear ruptures \cite{Passelegue2013} and precursory activity prior to stick-slip instabilities \cite{Passelegue2016}. Precursory activity to stick-slip instabilities has been also reported in experiments shearing a layer of granular material \cite{Riviere2018}. Other relevant results on similar experimental systems include remote triggering \cite{Johnson2008}, and the controlled slowing down of the dynamics \cite{Scuderi2016}. However, one common limitation of many of those laboratory experiments is the fact that they show a main dynamics consisting in a quasi-periodic stick-slip behavior with a narrow distribution of sizes, which do not correspond with the complex dynamics of real earthquakes described by the laws of seismicity. Other  experimental systems have also sheared a granular layer, aiming at mimicking the intermittent behavior of a tectonic fault \cite{Daniels2008, Walker2014}. Nevertheless it has been difficult to obtain a distribution of events that resembles the Gutenberg-Richter law \cite{Gutenberg1956} due to insufficient statistical sets of data.

Here we introduce a shear experiment capable of quantitatively reproducing the main statistical laws describing seismicity (Gutenberg- Richter law \cite{Gutenberg1956}, Omori law \cite{Omori1894}, distribution of inter-event times \cite{Corral2003}), as well as sharing many other qualitatively similarities with earthquake dynamics. As far as we know, it is the first time that such a quantitative agreement concerning simultaneously three main laws of seismicity is reported in a shear experiment (see more about the need of quantitative analogies at the Supp. Mat.). 
Its circular geometry allows the system to run continuously, capturing the considerable statistics required to analyze the dynamics of very large events, which rarely take place.

\textbf{Experimental System.} We study a 2D cylindrical pile confined in between two concentric fixed acrylic cylinders, and bounded by two rough circular rings (Fig.~\ref{fig:setup}a, b, Movie S1), with a dead load placed over the top ring and compressing the granular pile.
The top ring is free to move vertically but not to rotate, while the bottom one is slowly rotated with a period of $18.3\bar{3}~\text{hours}$, quasi-statically shearing the granular pile with a linear velocity of $48.84~\text{mm/hour}$ (approximately 12,600 times faster than the San Andreas fault, with an average slip rate of $33.9~\text{mm/year}$ \cite{Sieh1984}).
Thanks to a lever and a force sensor, we measure the torque $\Gamma(t)$ applied by the granular pile on the top ring. Six piezoelectric pinducers are inserted regularly in the top ring and simultaneously record acoustic emissions (AE).
Both measures are done at a rate of 100,000 samples per second.
The system is left to evolve for typical times of 24 hours.

During the shear, $\Gamma(t)$ shows an irregular stick-slip like behavior compatible with earthquakes dynamics, consisting in a continuous loading interrupted by intermittent drops with a large distribution of sizes (Fig.~\ref{fig:setup}c).
The detection of the torque drops (Fig.~\ref{fig:setup}d) is performed by applying a threshold to mechanical energy variations $\Delta(\Gamma^2(t))$ (see Appendix 1 for details).
Acoustic events are linked to local releases of energy taking place at the two-dimensional interfaces between grains \cite{Dubourg2017}.
The analysis of acoustic recordings (Fig.~\ref{fig:setup}e) is based on a Discrete Wavelet Transform (DWT) \cite{Chui1992} resulting in a time-frequency energy distribution map (Fig.~\ref{fig:setup}f), that is then processed to detect peaks, corresponding to the energy of the events (details in the Supp. Mat.).
The high number of events detected with both methods, respectively around 2,000 torque drops and more than 1.8 million acoustic emissions for a 24 hours experiment, allows to compute precise statistical characteristics of the system's behavior and to compare it with the dynamics of earthquakes.

\begin{figure}
\centering
\includegraphics[width=1.0\linewidth]{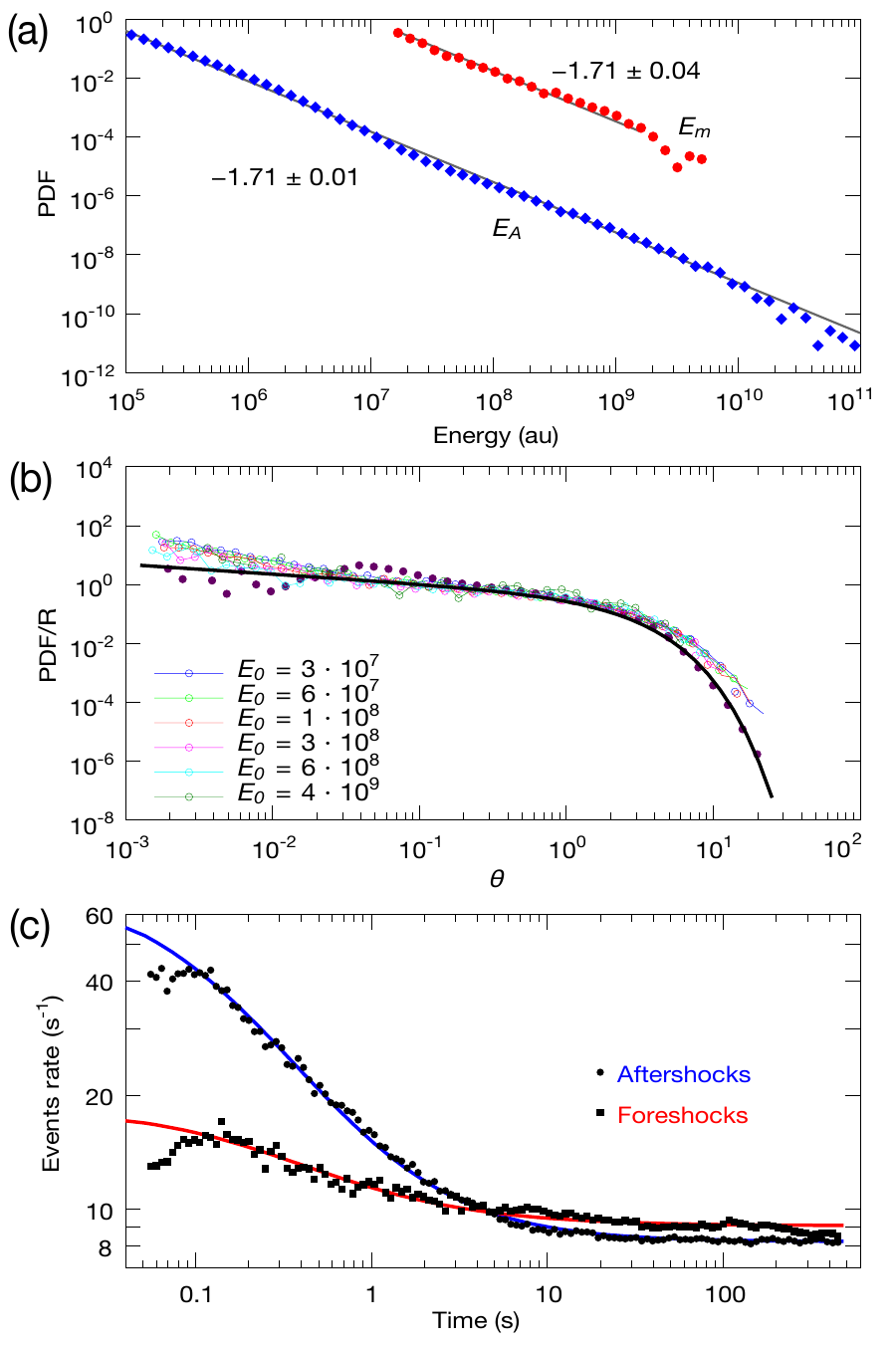}
\vspace{-0.5 cm}
	\caption{(color online) Quantitative reproduction of main seismicity laws.
	(a) Probability distribution of energies, either acoustic $E_a$ or mechanical $E_m$ for a 24-hours experiment.	
	Both distributions follow Gutenberg-Richter like laws $P(E)\sim E^{-\beta}$ with an exponent  $\beta=1.71$ (solid lines).
	(b) normalized probability distribution of $\theta=\tau_{E\geqslant E_0}/\tau^*_{E\geqslant E_0}$. Solid symbols: considering all the events. It follows the universal function $f(\theta)\sim\theta^{-0.3}\exp(-\theta/1.5)$ (solid line). Open symbols: considering different threshold values $E_0$.   
	(c) Average acoustic emissions rates respectively before (squares) and after (circles) a mainshock.
	They follow Omori laws, respectively $n(t_-)=9.05+2.78/(0.22 + t_-)^{0.8}$ for foreshocks and $n(t_+)=8.22+7.81/(0.12 + t_+)$ for aftershocks.}
	\label{fig:quakes_laws}
	\vspace{-0.6 cm}
\end{figure}

\textbf{Reproducing main seismicity laws.} 
\textbf{\textit{Gutenberg-Richter:}} On Fig.~\ref{fig:quakes_laws}a we present the probability distribution of both acoustic $E_\text{ac}$ and mechanical $E_\text{m}$ energies of detected events, on logarithmic intervals.
 The alignment of both types of energy on the abscissa axis has been obtained from Fig.~S1, where synchronous corresponding mechanical and acoustic events are represented.
For one given mechanical energy, the associated acoustic one is obtained as the median energy of corresponding acoustic emissions.

The two probability density functions behave like power laws $P(E)\sim E^{-\beta} $ with exponents $\beta_\text{m}=1.71 \pm 0.04$  and $\beta_\text{ac}=1.71 \pm 0.01$ for mechanical and acoustics energies respectively, obtained by a maximum likelihood method \cite{Clauset2009, Alstott2014}.
The AE energies spread over six decades while the mechanical energies cover only three decades, showing the better sensitivity of the acoustic detection.
This power-law behavior is to be compared with the Gutenberg-Richter law \cite{Hanks1979, Choy1995} which states that the PDF of radiated energies of globally measured real earthquakes follows a power-law with an exponent $\beta=5/3 = 1.67$. 

\textbf{\textit{Inter-event times:}}
By defining a threshold in energy $E_0$ (or in magnitude $M=2/3~\log E -2.9$, which is more used in earthquakes studies \cite{Choy1995}) we can analyze the inter-event time between two consecutive events  $\tau_{E\geqslant E_0}$ (or $\tau_{M\geqslant M_0}$). 
In nature, the rate of seismicity $R_{M\geqslant M_0}$, defined as the number of earthquakes larger that a given magnitude $M_0$ per unit time varies depending on the region (eg. $R_{M\geqslant2}\sim15,000$ earthquakes/year for California \cite{Felzer2008}).
However, the distribution of $\theta=\tau_{M\geqslant M_0}/\tau^*_{M\geqslant M_0}$  is a universal function following $f(\theta)\sim\theta^{-0.3}\exp(-\theta/1.5)$ for all seismic zones \cite{Corral2003}, where $\tau^*_{M\geqslant M_0}$ is a characteristic time defined as the inverse of the rate of seismicity $\tau^*_{M\geqslant M_0} = 1/R_{M\geqslant M_0}$.
Although the rate of $all$ our AE events corresponds to $R=17.35$ events/s (more than 36,000 times higher than Californian earthquakes), which gives a characteristic time $\tau^*=57.64$ ms, the $\theta$ distribution of our AE events follows quantitatively the same universal function (Fig.~\ref{fig:quakes_laws}b). 
In the case where all the events are considered, $\theta =10^{-1}$ corresponds to a $\tau = 5.764$ ms, which corresponds to the duration of the events (see Fig.~\ref{fig:setup}e).
 This sets a limit to the inter-event time distributions and we can notice a deviation of the distribution from the power-law behavior for values of $\theta <10^{-1}$ (Fig.~\ref{fig:quakes_laws}b).

The universal function tells us that the system ``remember'' the events of energy $E\geqslant E_0$ during a time corresponding to their characteristic time $1/R_{E\geqslant E_0}$, where the distribution is a power law.
However, this memory, as in the case of earthquakes, is quite weak an carries no predictive capabilities \cite{Touati2009}.
For longer inter-event times the exponential tail of the distributions indicates that the events are independent. 

For large threshold values ($E_0$), the distributions deviate from the universal law both for small and large $\theta$ values (Fig.~\ref{fig:quakes_laws}b).
The increase of short inter-event times is a direct consequence of the increase of the activity associated to aftershocks and foreshocks in the dynamics, which is also a feature of real earthquake data \cite{Corral2003}. 
The increase of long inter-event times is linked to insufficient statistics. 
In order to verify that, we have analyzed the distribution of inter-event times $\tau_{\ell}$ for large events only ($E_0 = 10^8$) (Fig.~S2a).
It presents two regimes: a clustering of events for short inter-event times ($\tau_{\ell}<\tau^*$), and an exponential decay  $D(\tau_{\ell})\sim\exp\left(-\tau_{\ell}/\tau_c\right)$ indicating that large events separated by long inter-event times ($\tau_{\ell}>\tau^*$) are independent and follow a Poissonian process.
When we increase the energy threshold $E_{0}$ defining the large events (notice that $R_{E\geqslant E_{0}}\sim E_{0}^{1-\beta}$), we expect a linear relation between $\tau_c$ and $\tau^*$.
However, we find that the increase of $\tau_c$ is slower than linear, and the best fit shows a power law with an exponent 0.86 $\pm$ 0.01 (Fig.~S2b).
This deviation may be caused by a lack of statistics concerning very large events and eventually may be used as an analytical tool to estimate biases in the obtained results due to insufficient statistical sets of data.


\textbf{\textit{Omori:}} By defining the large acoustic emissions ($E_\text{ac}\geqslant 10^8$) as mainshocks, corresponding to about 4,500 events, we are able to reveal the existence of foreshocks and aftershocks following Omori laws as for real earthquakes \cite{Omori1894}: 
$n(t)=A/(c+t)^p+B$, where $A/c^p$ gives the rate increase associated to the mainshock, $B$ the background rate of the earthquakes, $c$ the time offset (positive and close to zero) \cite{Enescu2007} and $p$ the Omori exponent, around 1.
On Fig.~\ref{fig:quakes_laws}c we show the average of the AE rate of foreshocks and aftershocks around the mainshocks (where time of foreshocks is $t_-=t_m-t $ and time of aftershocks $t_+=t-t_m$, with $t_m$ the time of the mainshock).
The foreshocks rate follows an Omori-like increase $n(t_-)=9.05~\text s^{-1}+2.78/(0.22~\text s + t_-)^{0.8}$ with a reduction of the activity in the last 0.1 s preceding the mainshock. 
Just after the mainshock, the aftershocks rate presents first a plateau, associated in real earthquakes to catalogues incompleteness (due to large quakes masking smaller ones), followed by a power-law decrease as $n(t_+)=8.22~\text s^{-1} +7.81/(0.12~\text s + t_+)^{1.0}$. 
When choosing larger threshold values defining the mainshocks (Fig.~S2b) the number of events reduces; but the behavior is qualitatively the same. 


 

 \begin{figure}[b!]
\centering
\vspace{-0.5 cm}
\includegraphics[width=1.0\linewidth]{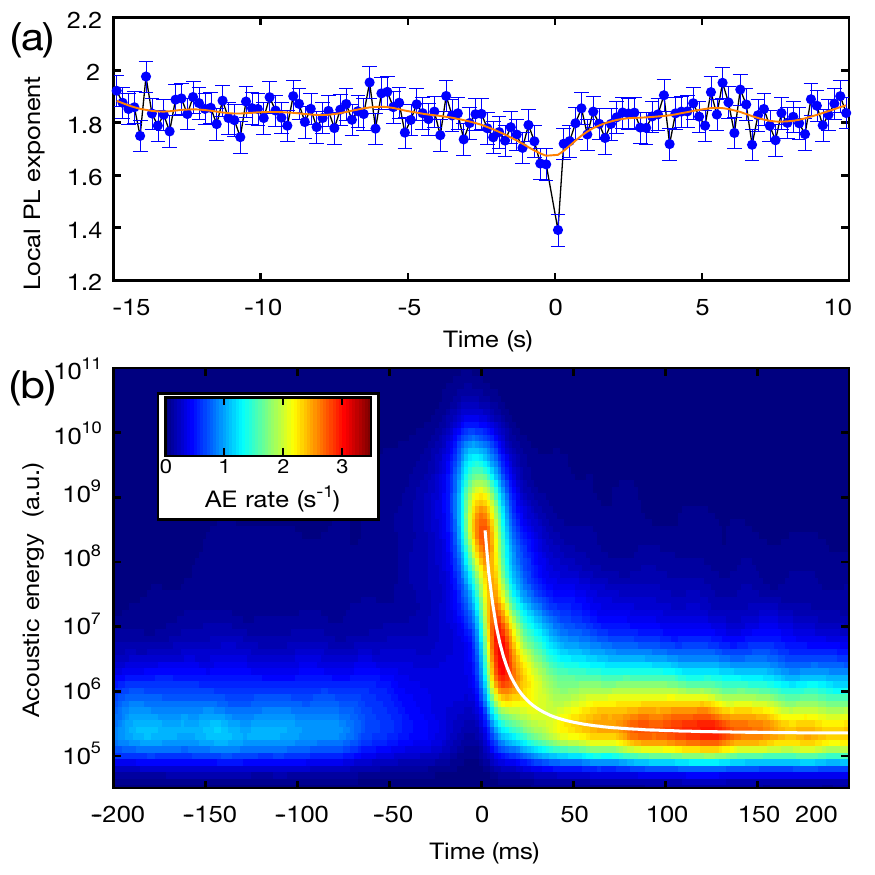}
\vspace{-0.5 cm}
	\caption{(color online) Around mainshocks.
	(a) Local power-law exponent computed on all acoustic foreshocks or aftershocks detected in fixed 200~ms-windows centered at a given time from the mainshock.
	(b) Histogram of acoustic emissions energies on 2 ms intervals during $\pm$200 ms around a large acoustic emission.
	The AE rate is displayed as shades of color, with a corresponding numerical value averaged over the 4,500 mainshocks.
	We observe an abundance of high amplitudes right after the mainshock, with a decrease of the logarithm of the most probable acoustic energy value as a power-law of time (with an offset) $5.33+\left[60.34/(t+44.91)\right]^{4.82}$, also displayed as a white solid line.
	}
	\label{fig:stat}
\end{figure}

\textbf{Around mainshocks.} Beyond analyzing the main laws of seismicity, we can also study the system's behavior around mainshocks.
We divide the time axis into windows of 200 ms duration, from 15 seconds before to 10 seconds after a mainshock, allowing computing a local probability distribution of acoustic energies found at a given time of any of the large emission.
We find power law distributions, with a variable exponent represented on Fig.~\ref{fig:stat}a.
Far from a mainshock, we find a constant value close to 1.85, slightly bigger than the global $\beta$-value of 1.71.
This difference is caused by the  selection of time windows without extreme events since they are used as reference mainshocks.
For about 4 seconds before the mainshock, the exponent shows a continuous but slight decrease that accelerates in the last second to reach a value of about 1.6 just before the mainshock.
The first calculated $\beta$-value after the mainshock reaches 1.4 and then it jumps again to 1.85. 
These low $\beta$-values indicate an abundance of high-energy events, which are independent of the increase of the rate of the events (foreshocks and aftershocks).
This subject is a source of controversy in the Statistical Seismology community \cite{Davidsen2011, Lippiello2012}. 
Indeed, the analysis of local and relative fast variations of the $\beta$-value in real earthquakes is well documented, often associated to correlation between magnitudes \cite{Lippiello2012}, but always affected by both the intrinsic lack of statistics and the incompleteness  of the catalogues \cite{Davidsen2011}. 
As our experiments rely on single measurements captured during a relative short period (one or a few days), the reported results are less affected from catalogue incompleteness than real earthquake data, making our statistical results very reliable.


We can focus on the two closest data points around the mainshock of Fig.~\ref{fig:stat}a. 
The corresponding 2D histogram (Fig.~\ref{fig:stat}b) shows that high-energy events cluster just after the mainshock. 
They have been detected --thanks to the wavelet analysis-- inside the acoustic envelopes of the mainshocks, which can last for up to 30 ms. 
As large events correspond to rearrangements involving a large number of grains, it is expected to detect secondary peaks associated to the mainshocks.
A priori the Omori law may not explain the statistics during these dynamic rearrangement, thus magnitude correlations seems the appropriated term to refer to this clustering of high energy events. 
Indeed, the activity seems to restart progressively from 40 ms having a maximum around 120 ms. 
These results, coherent with the aftershocks displayed on Fig.~\ref{fig:quakes_laws}c, indicate that the flat part of the Omori law between 50 ms and 120 is not provoked by missing small events masked by previous large ones.
We can also notice a clear quiescence in the 60 ms interval preceding the mainshock, also coherent with the decrease of the foreshock activity close to the mainshocks (Fig.~\ref{fig:quakes_laws}c), a phenomenon that has been often reported prior to very large earthquakes \cite{Kanamori1981}. 
Longterm decrease of $\beta$-values has also been reported preceding very large earthquakes \cite{Nanjo2012}, and also in controlled experiments \cite{Sammonds1992} and simulations \cite{Kun2013}.

\textbf{Discussions.}
A very large distribution of thresholds, required to achieve an earthquake-like dynamics, is directly related to the heterogeneous character of the system.
In subcritical fracture experiments like those cited in the introduction  \cite{Scholz1968, Maloy2006, Baro2013, Stojanova2014, Bares2018}, the heterogeneity is provided by a structural disorder, and the competition between the advancing crack and the fracture thresholds may result in a Gutenberg-Richter-like distribution of event sizes. 
 However, most shear experiments present a main dynamics composed of quasi-periodic stick-slip events with a narrow distribution of sizes, which may be a consequence of a lack of disorder. 
 In the case of similar experiments shearing a granular layer \cite{Riviere2018, Johnson2008, Scuderi2016}, a very high number of particles and three-dimensional force chains may be responsible for an ``averaging'' effect that reduces the heterogeneity of the system, resulting in a regular stick-slip dynamics similar to the one obtained in solid flat interfaces.
 
 In our system, the granular force network  \cite{Miller1996, Howell1999, Majmudar2005} provides an evolving heterogeneity in terms of energy thresholds that is the key ingredient of the dynamics, and it is responsible for a distribution of events that resembles the Gutenberg-Richter law \cite{Gutenberg1956, Choy1995}. 
 The structure, dynamics and sizes of these heterogeneities in an actual fault remain as open questions. 
 However, the two-dimensional nature of both our system and the inter-grain friction, and the low dimensionality of the force network (which may depend on the pressure  between the plates \cite{Lherminier2014}) may serve as hints to eventually find them (see more details in Supp. Mat.).
 Our system is able to reproduce quantitatively the main statistical laws of seismicity, which indicates that both earthquakes and our experiment are governed by a similar physics, and opens a new pathway to the investigation of earthquake-like dynamics at a laboratory scale.

\begin{acknowledgments}
We thank E. Altshuler, T. Bodin, T. Hatano, C. Lasserre and M. Métois for very useful discussions.
We acknowledge support from the AXA Research Fund. O. R. acknowledges support from the Chaire-CNRS program and the University of Tokyo for the invited researcher position at the Earthquake Research Institute. This work was partly supported by the Research Council of Norway through its centers of Excellence funding scheme, project number 262644.
\end{acknowledgments} 

\bibliography{paperbib}

\end{document}


\setlength{\baselineskip}{0.4cm} \addtolength{\topmargin}{1.5cm}

\title{Revealing the structure of a granular medium through ballistic sound propagation \\SUPPLEMENTAL MATERIAL}

\title{Continuously Sheared Granular Matter Reproduces in Detail Seismicity Laws \\SUPPLEMENTAL MATERIAL}

\author{S. Lherminier}
 \affiliation{Institut Lumi\`ere Mati\`ere, UMR5306 Universit\'e Lyon 1-CNRS, Universit\'e de Lyon 69622 Villeurbanne, France.}
 
\author{R. Planet}
 \altaffiliation{Present address: Departament de F\'isica de la Mat\`eria Condensada, Universitat de Barcelona, Mart\'i i Franqu\`es 1, E-08028 Barcelona, Spain and Universitat de Barcelona, Institute of Complex Systems, Mart\'i i Franqu\`es 1, E-08028 Barcelona, Spain.}
 \affiliation{Institut Lumi\`ere Mati\`ere, UMR5306 Universit\'e Lyon 1-CNRS, Universit\'e de Lyon 69622 Villeurbanne, France.}
 
 \author{V. Levy dit Vehel}
 \affiliation{Institut Lumi\`ere Mati\`ere, UMR5306 Universit\'e Lyon 1-CNRS, Universit\'e de Lyon 69622 Villeurbanne, France.}
  
\author{G. Simon}
 \affiliation{Institut Lumi\`ere Mati\`ere, UMR5306 Universit\'e Lyon 1-CNRS, Universit\'e de Lyon 69622 Villeurbanne, France.}
 
\author{L. Vanel}
 \affiliation{Institut Lumi\`ere Mati\`ere, UMR5306 Universit\'e Lyon 1-CNRS, Universit\'e de Lyon 69622 Villeurbanne, France.}
 
\author{K. J. M\aa{}l\o{}y}
 \affiliation{PoreLab,  Department of Physics, University of Oslo, P. O. Box 1048, 0316 Oslo, Norway.}
 
\author{O. Ramos}
\email{osvanny.ramos@univ-lyon1.fr}
 \affiliation{Institut Lumi\`ere Mati\`ere, UMR5306 Universit\'e Lyon 1-CNRS, Universit\'e de Lyon 69622 Villeurbanne, France.}

\date{\today}

\maketitle


 \section{Methods}\label{AppMethods}
 
\subsection{1. Experimental setup}
The experimental setup consists of two fixed, transparent, and concentric acrylic cylinders with respective diameters of 28~cm (inner) and 29~cm (outer), so that a monolayer of approximately 3500 disks can be introduced into the 5 mm gap (Fig.~1a, b, Movie S1).
These disks, of 4~mm thickness and 6.4~mm and 7.0~mm diameter (in equal proportion) to avoid crystallization, have been 3D-printed in \textit{Durus White 430} thanks to a \textit{Objet30} printer.
Two rings -- 3D-printed within the same material as the grains and consisting of 99 half cylinders (of diameter $d =$~6.4~mm) separated by $\sqrt{2}d \approx$~9~mm each -- are introduced respectively at the bottom and on top of the disks monolayer and constrain the pack using a dead load of 276~N.
The load was also varied between 126~N and 326~N in order to analyze its influence on the energy distribution of the events (Fig.~S3).
The translucent and photoelastic properties of the \textit{Durus} material allows the visualization of the stress inside the disks when placing the experimental setup between two circular polarizers (see close-up in Fig.~1b).
The top ring is free to move vertically but a torque-meter blocks its rotation. 
The bottom ring is fastened through a roller chain to a gear mechanism that reduces 2200 times the rotation of a stepping motor turning with a period of 30 s. Consequently, the bottom ring rotates at a constant and very low speed with a period of $18.3\bar{3}~\text{hours}$, quasi-statically shearing the granular pile with a linear velocity $v =$~48.84~mm/hour. A compressed spring of stiffness $K_\text{gear} =$~230~kN/m holds the chain under tension and defines the stiffness of the apparatus.
As a result, shear stresses build up on the packed disks.
The release of the stress happens with sudden avalanches, \emph{i.e.} reorganizations of the packing, associated with acoustic emissions.
A clear shear band is created separating two parts of the granular medium (see movie S1), one reacting to the movement of the bottom ring, and another that barely moves \cite{Henann2013, Kuwano2013}.
The zero-frequency Young modulus of \textit{Durus} material is $Y \approx$~ 100~MPa, which contrasts with classical experiments using photoelastic disks with a Young modulus $Y=$~4~MPa \cite{Daniels2008, Majmudar2005}.
Our grains can hold a much larger stress without a considerable deformation, which favours both the acoustic propagation and image analysis \cite{Lherminier2014}.

\subsection{System monitoring}
To characterize the mechanical response of the sheared material we extract the resisting torque of the system using a steel lever fastened to the upper ring and a force sensor \textit{Interface SML-900N} (of range 900~N and stiffness 1.1$\times10^7$~N/m, both playing the role of a torque-meter.
Acoustic emissions are recorded using 6 piezoelectric pinducers (\textit{VP-1.5} from \textit{CTS Valpey Corp.}) inserted in regularly-placed adjusted holes in the upper ring.
A constant acoustic coupling is ensured using silicon oil at the edge of the pinducers.
The acoustic and mechanical signals have been recorded using a \textit{NI-USB-6366} card.
The acoustic frequencies generated by the system have been measured using a sampling rate of 2~MHz and were found to be lower than 25~kHz, so the sampling rate has been chosen at 100~kHz to optimize the recording (typically 24 hours).

\subsection{Torque signal analysis}
The torque signal (applied by the granular pile on the fixed boundary) is also recorded at 100~kHz.
Because of inertia, torque variations are slow so the signal can be decimated at 1~kHz.
The signal is squared and differentiated to obtain the variations of mechanical energy.
Indeed, by considering the system as a series of springs, one corresponding to the gear mechanism with stiffness $K_\text{gear}$, one to the grains pile $K_\text{grains}$ and one to the force sensor $K_\text{sensor}$, we can compute the energy associated with a torque drop as
$E_\text{m} = \left( {\Gamma_i}^2 - {\Gamma_f}^2 \right)/2K_\text{tot}$
with ${K_\text{tot}}^{-1} = {K_\text{gear}}^{-1} + {K_\text{grains}}^{-1} + {K_\text{sensor}}^{-1} \approx {K_\text{grains}}^{-1}$ and $\Gamma_i$ (respectively $\Gamma_f$) is the initial torque at the beginning of the torque drop (resp. final torque at the end of the drop).
The value of $K_\text{tot}\simeq~K_\text{grains}$ can be estimated from the slope of torque signal during loading periods on Fig.~1c and is found equal to $K_\text{tot}=\Delta F/ \Delta s = (\Delta\Gamma/r)/(v \Delta\ t) \approx$ 1,724~N/m.
As we are interested in
relative values, we will compute the mechanical energy as $E_\text{m} \sim ({\Gamma_i}^2 - {\Gamma_f}^2)$.

This temporal series of energy variations is filtered by a Butterworth 6th-order low-pass digital filter with a cut-off frequency of 50 Hz.
An avalanche will manifest itself on the signal by a short peak of energy variation, corresponding to the dissipation of energy during the avalanche.
Once detected a peak top, its beginning (respectively end) is found by applying a threshold over the noise standard deviation in the few milliseconds preceding (resp. following) the peak top.
Amplitude of the corresponding avalanche is obtained by integrating the mechanical energy variations over the peak length.

\subsection*{4. Acoustic signal analysis}
The acoustic signal is recorded at 100~kHz after a 1st-order low-pass filtering at 100 kHz and an amplification by 17.6 dB thanks to a locally designed amplifier.
The analysis is based on a Discrete Wavelet Transform (DWT), \cite{Chui1992} operated on 128 frequencies logarithmically distributed between 1 and 20 kHz and on instants separated by 50 $\mu$s.
The DWT is computed thanks to the Time-Frequency ToolBox of Matlab.
The result of the DWT is a scalogram, which is a 2D map of energy of the signal versus time and frequency.
The logarithm of the scalogram is smoothed by a Gaussian blurring scaling on 12 frequency bins and 0.6 ms.
Peaks are detected, with their corresponding times, frequencies and energy values.

\subsection{Statistic laws}
After the beginning of the experiment, the force increases during around half an hour to reach a stationary situation. The avalanche behavior on this transient zone is excluded from the statistical analysis. 

For the Gutenberg-Richter law (Fig.~2a), the probability distribution of amplitudes has been computed on logarithmic bins (10 bins per decade) distributed between the minimum and maximum amplitude of detected avalanches, either for mechanical energy or acoustic energy.

For each interval of bins, the number of events with an amplitude belonging to the interval is divided by the width of the interval.
The corresponding exponent has been obtained by a maximum likelihood method \cite{Alstott2014}.

For the Omori law (Fig.~2c), the acoustic avalanches rate has been computed in average around large energy events ($E_\text{ac}\geqslant 10^8$).
Again the time bins have been defined logarithmically and the events rate has been normalized by intervals widths.
The fit has been computed by the non-linear least-squares Marquardt-Levenberg algorithm. 
For the inter-event times distribution (Fig.~2b), they have been computed as difference between times of two successive avalanches.
In the case of the whole series of avalanches, the inter-event times have been computed without taking into account the amplitudes of avalanches.


The 2D histogram of acoustic energies versus time (Fig.~3) has been computed by dividing time in 2 ms intervals from 200 ms before to 200 ms after the mainshock detection, and by dividing acoustic energies range into 66 energy bins (10 per decade).
A blur is applied on the figure under the form of a Gaussian filter scaling on 2.5 bins in each direction.

\newpage
 \section{Supplementary Figures}
 
\begin{figure}[h!]
\includegraphics[width=0.5\linewidth]{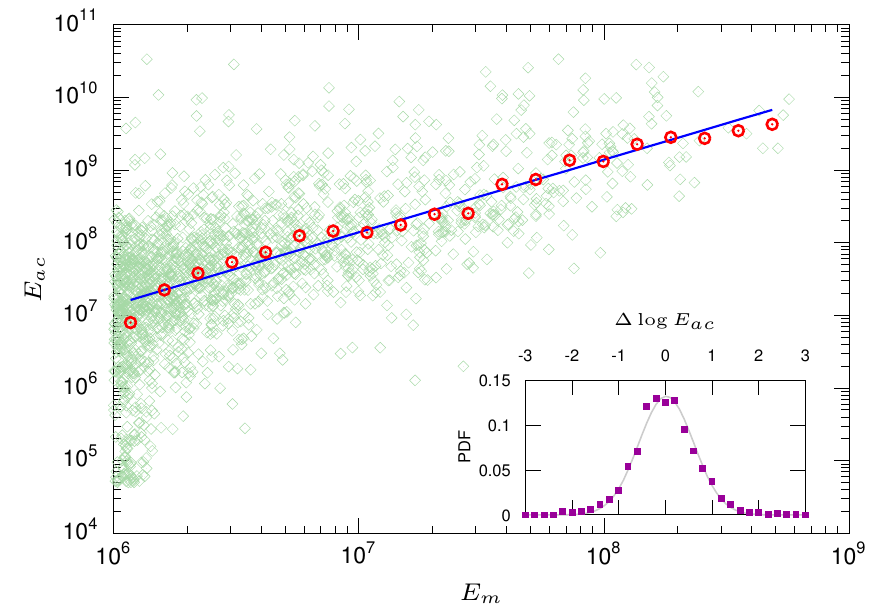}
\caption{Equivalence of mechanical and acoustic monitoring.
	For each torque drop, we display the energy of the main acoustic emission associated versus the mechanical energy (green diamonds).
	By dividing the range of mechanical energies into 20 intervals, we compute the median acoustic energy for torque drops that belong to each interval (red circles)
	We find a linear dependence displayed as a solid line.
	Inset: distribution of acoustic energies logarithms deviations with respect to their median value, averaged for the 20 mechanical energy intervals and fitted to a normal distribution.
	\label{fig:amplitudes_all}
	}
\end{figure} 

 \begin{figure}[h!]
\centering
\includegraphics[width=0.5\linewidth]{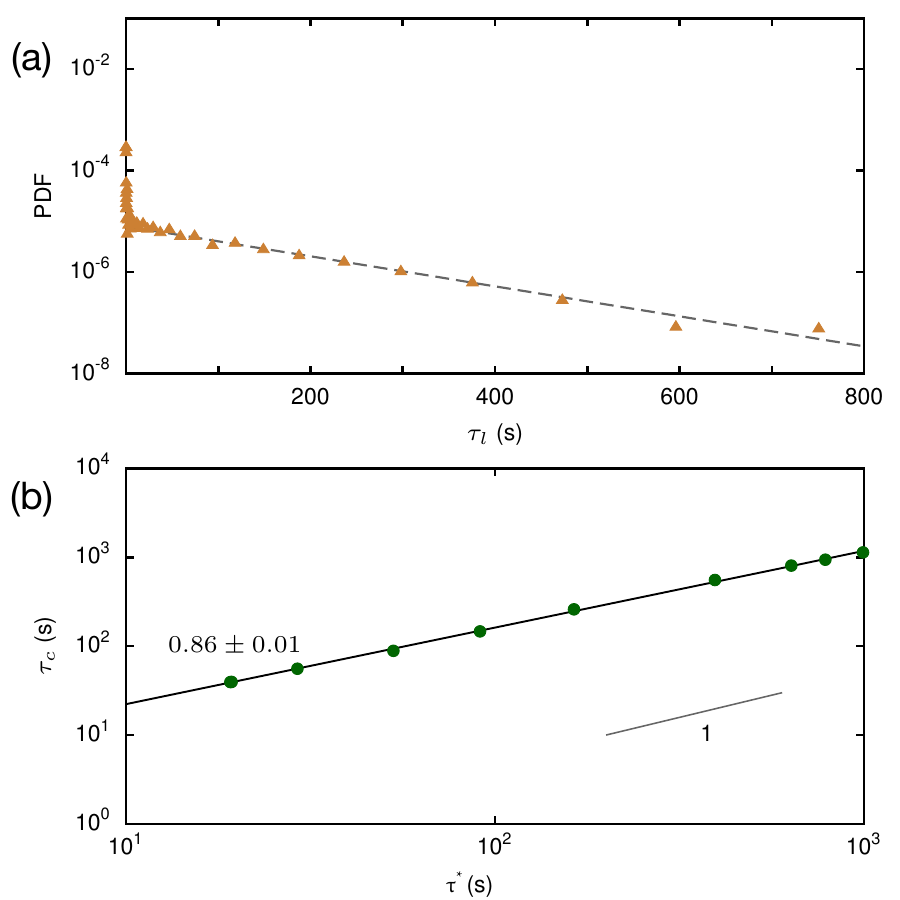}
	\caption{Exponential tails of inter-event time distributions.
	(a) Probability distribution of inter-event times $\tau_{\ell}$ for large events only ($E_\text{ac}\geqslant 10^8$). It shows an exponential decay for long inter-event times $D(\tau_{\ell})\sim\exp\left(-\tau_{\ell}/\tau_c\right)$, with $\tau_c=135~\text{s}$ (dashed line).
		(b) When the energy threshold defining a mainshock $E_0$ is continuously  increased, the $\tau_c$ values follow a power law with an exponent 0.86 (solid line) with respect to the inverse of the rates of large events. A thin line indicates a linear relation. }
	\label{fig:longWT}

\end{figure}

\begin{figure}
	\centering
	\includegraphics[width=0.5\linewidth]{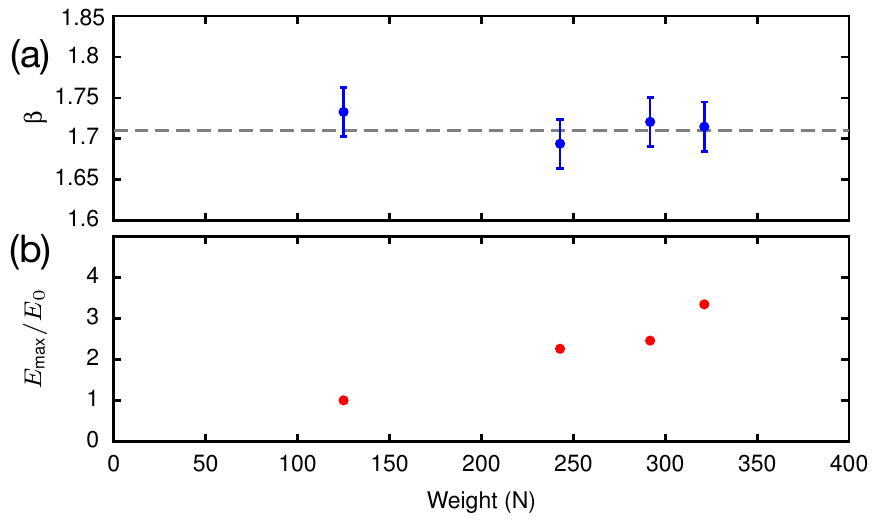}
	\caption{Effect of dead load variation on avalanches statistics (measured as mechanical energy drops).
	(a) The exponent of the distribution of events' energies (equivalent to the Gutenberg-Richter law) seems robust when the dead load (simulating the force between the tectonic plates) is varied. (b)  The energy of the largest avalanches, normalized to the energy of the largest avalanche for the minimum applied load, increased with the applied load.
	\label{fig:weight_var}
	}
\end{figure}

\section{Insights into real earthquakes}
Our system is able to reproduce quantitatively the main statistical laws of seismicity, which indicates that both earthquakes and our experiment are governed by a similar physics, and opens a new pathway to the investigation of earthquake-like dynamics at a laboratory scale.
 Although our apparatus is composed of two plates that compress and shear a granular material, it is important to point out that this system replicates the dynamics of a tectonic fault, but not the fault itself. 
 Our grains do not necessarily represent the granular material inside a real fault. 
 Our granular system provides, thanks to the force network \cite{Miller1996, Howell1999, Majmudar2005}, a very heterogeneous and evolving matrix to store the mechanical energy generated by the relative movement of the plates. 
 This evolving heterogeneity in terms of energy thresholds is the key ingredient of our system, and it is responsible for a distribution of events that resembles the Gutenberg-Richter law \cite{Gutenberg1956, Choy1995}. 
 
In subcritical fracture experiments like those cited in the introduction  \cite{Scholz1968, Maloy2006, Baro2013, Stojanova2014, Bares2018}, the heterogeneity is provided by a structural disorder, and the competition between the advancing crack and the fracture thresholds may result in a Gutenberg-Richter-like distribution of event sizes. 
 However, most shear experiments between two solid blocks present a main dynamics composed of quasi-periodic stick-slip events with a narrow distribution of sizes, which may be a consequence of a lack of disorder. 
 In the case of similar experiments shearing a granular layer \cite{Riviere2018, Johnson2008, Scuderi2016}, a very high number of particles and three-dimensional force chains may be responsible for an ``averaging'' effect that reduces the heterogeneity of the system, resulting in a regular stick-slip dynamics similar to the one obtained in solid flat interfaces.
 The structure, dynamics and sizes of these heterogeneities in nature remain as open questions. 
 
In the theories of critical phenomena it is known that the exponent values describing the dynamics depend heavily on the system dimension \cite{Stanley1987}. 
We have obtained very similar exponent values than the laws of seismicity in a two-dimensional system. 
In addition, the main physical process corresponds to the inter-grains friction, which is also a two-dimensional problem, and the fractal dimension of the force chains has values smaller than two and may depend on the pressure between the plates \cite{Lherminier2014} and on the nature of the friction.
These structural hints may be useful in the search of the heterogeneities responsible for the Gutenberg-Richter law in nature.

  \begin{figure}[t!]
	\centering
	\includegraphics[width=11.4cm]{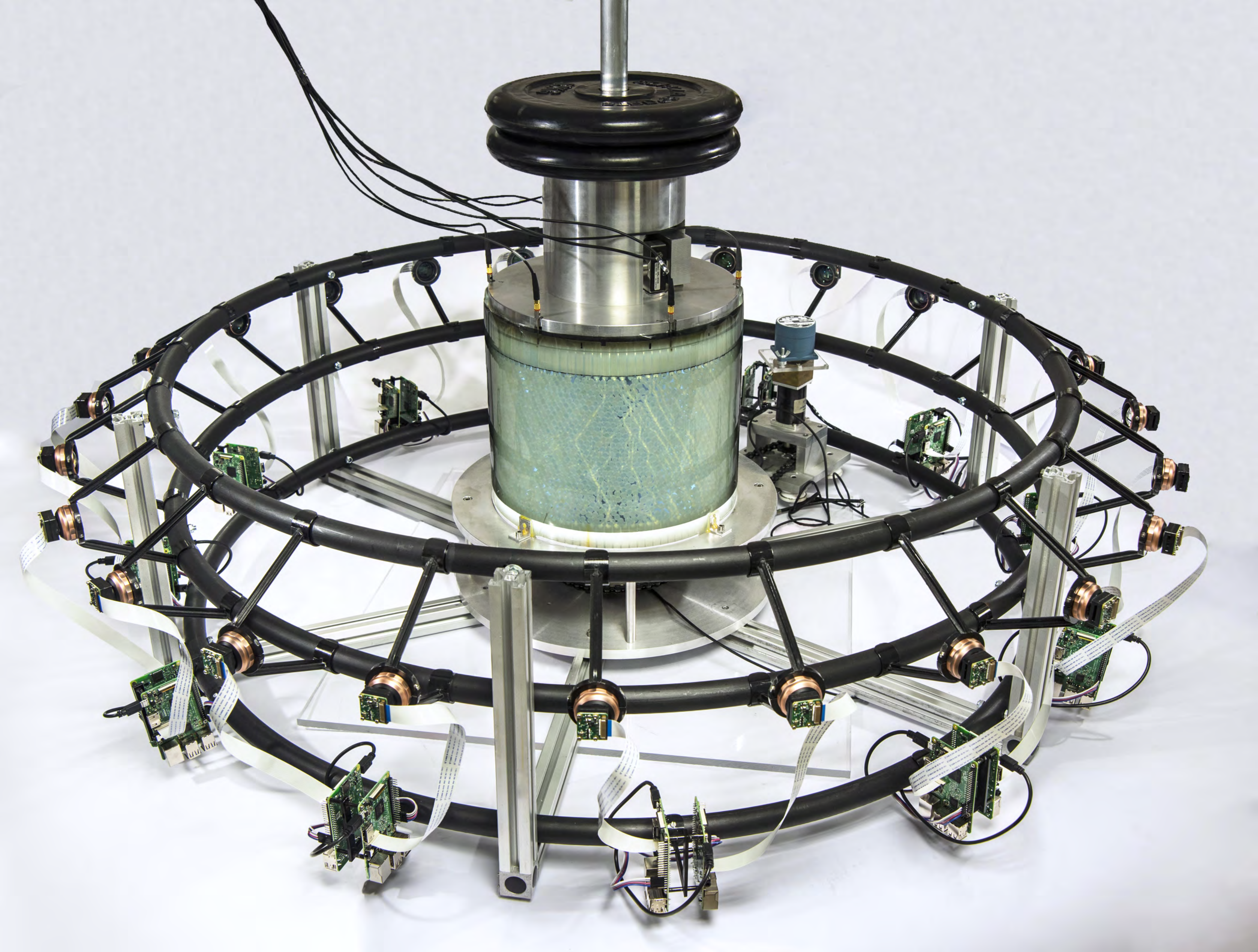}
	\caption{Imaging setup.
	Simplified image of the experimental system containing the imaging setup. It consists of 24 independently controlled raspberry-Pi cameras that take images of the granular medium with a frequency of 0.5 Hz.
	\label{cameras}
	}
\end{figure}
Compared to the huge complexity of a tectonic fault, our system is extremely simplified.
This is an important asset, because it helps understanding the essential ingredients of the global dynamics, but it is also a limitation, in particular because several key elements of the fault dynamics are absent (or hidden) in our system.
A clear example is fluid pressurization \cite{Yamashita2018}, which is a major ingredient into the triggering dynamics affecting the local friction both due to lubrication and normal force variations.
 Our experiment (with no fluids at all) is capable to reproduce the main laws of seismicity.
  Which is then the role of the fluids into the generation of these laws in the reality of a fault? 
 Concerning Gutenberg-Richter, our results indicate the requirement of a very heterogeneous matrix where the energy is stored, creating a very large distribution of energy thresholds. 
 Could the water network inside the ground be responsible of this heterogeneity, playing a similar role to the force network in our granular structure?  
 We do not know the answer. 
 However, this example illustrates one goal of our simplified system: providing a set of key ingredients that serve as a guide to find the actual ones in the reality of a fault. 
 We are, therefore, complementary to the other experimental systems focusing their main efforts on single events.  
 
Modifying the main parameters (confining pressure, disorder, dissipation, etc.) in order to better understand their roles in the dynamics may allow explaining the causes and consequences (i.e., possibility of prediction \cite{Main1996}) of the phenomenological laws describing seismicity. 
 We have preliminary results indicating a robustness of the exponent value ($\beta$-value) of the Gutenberg-Richter-like energy distribution with the applied load (Fig.~S3). 
 However, we expect that a further increase of the load will result in a reduction of the $\beta$-value \cite{Hatano2015} due to modifications of the force network \cite{Lherminier2014}. 
 We have recently installed 24 cameras (fig.~\ref{cameras}) required to analyze the local structure of our system (position of the grains and force network). 
 These results will bring valuable information on how the evolution of the structure relates to the dynamics of the system.

 \section{The need of quantitative analogies}

In physics there is a habit to report power law distributions of event sizes or energies (often regardless their exponent values) as analogues of the Gutenberg-Richter law. 
In this section we focus on the implications of the power law exponent of the Gutenberg-Richter law and, as a consequence, the need of quantitative analogies in the quest of building reliable models of earthquakes. 
 
\begin{figure} [t!]
\centerline{\includegraphics[width=0.7\textwidth]{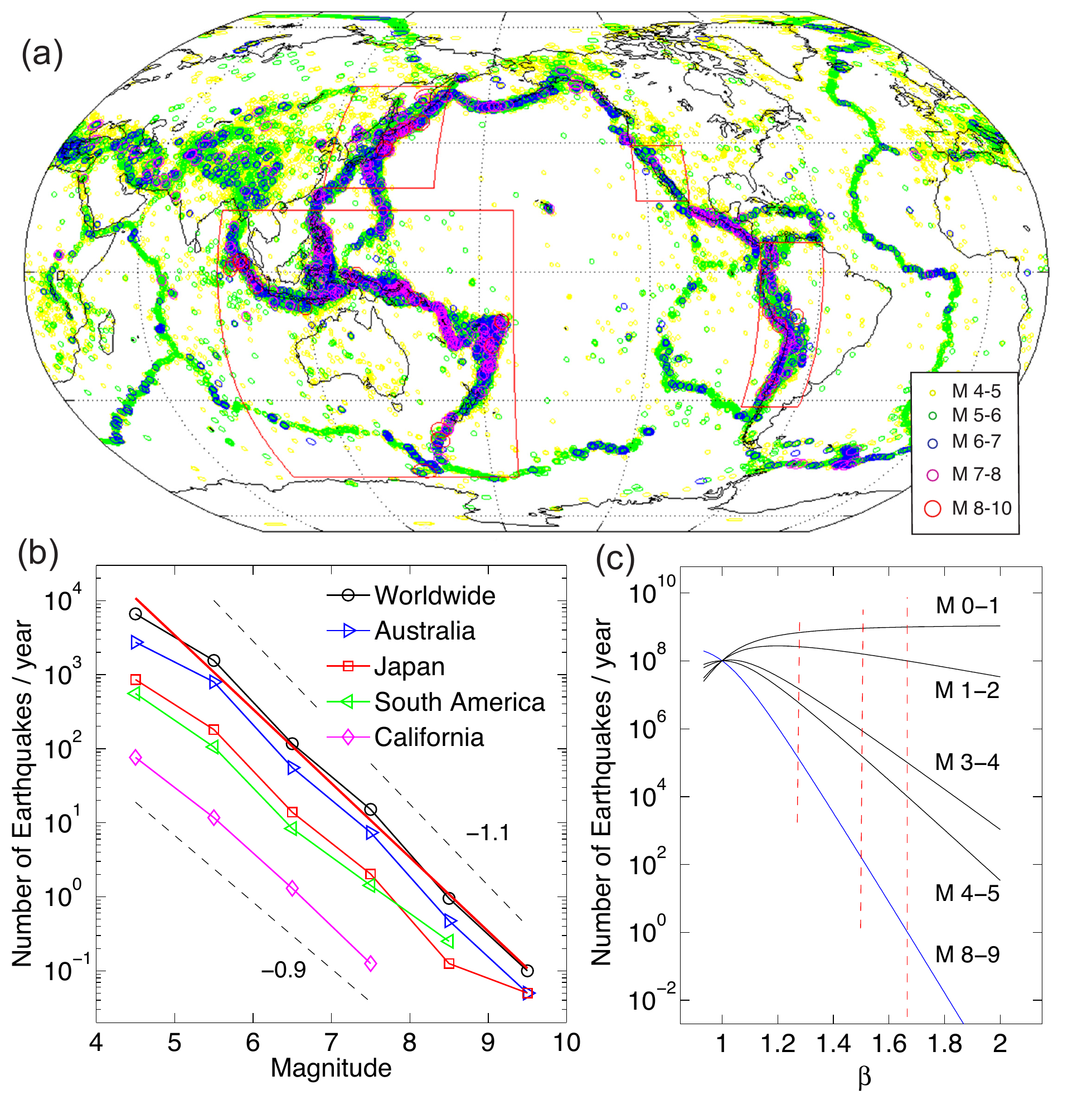}}
\caption{Distributions of earthquakes. (a) Earthquake world map, with four selected zones with different sizes and activities. (b) Distributions of earthquake's magnitude worldwide and in each selected zone. The red line, with an slope $-1$ is the best fit, while the dashed lines indicate that the error of the slope is around $\pm 0.1$. (c) Number of earthquakes of a given magnitude per year considering different exponent values. The vertical lines indicate the critical value of the exponent for the OFC model of earthquakes ($\beta=1.27$), the value of the exponent for mean-field models of avalanches ($\beta=3/2$) and the exponent for real earthquakes ($\beta=5/3=1.67$). }\label{Quakes}
\end{figure}

The fig.~\ref{Quakes}a shows the statistics of earthquake occurrence in the whole planet and in four zones with different sizes and activities. 
The data has been extracted from the U.S. Geological Survey (USGS) \cite{USGS}.
It covers a 40 years period, from 1973-01-01 00:00:00 to 2013-12-31 23:59:59.
In order to guarantee the completitude of the catalog, only the events with magnitude $M \geqslant 4.0$ have been considered.

We want to focus on the differences between small events and large ones, so we need to know, for example, the number of earthquakes/year with magnitude 4.0 $\leqslant M<$ 5.0, instead of the classical analysis considering the number of earthquakes/year larger than a given magnitude $M>$ 4.0.
 Therefore, instead of analyzing the classical cumulative form of the Gutenberg-Richter law: $log [P(m>M)] =a - b M$, we will focus on the form: $log [P_{int}(M)] =a_2 - b M$ (Fig.~\ref{Quakes}b), where $M$ corresponds to the magnitude and $P_{int}$ indicates that the measurements involve the integration of the values in a given interval of magnitudes. Both distributions share the same b-value in the case of $b\sim1$.
 
The solid red line and the dashed ones in Fig.~\ref{Quakes}b show that the b-value $b=1 \pm 0.1$ (provided a large statistics). 
By substituting $M$ into the definition $M \equiv 2/3~log (E) + constant$, where E corresponds to the energy released by a quake \cite{Hanks1979}, one gets $P_{int}(E) \sim E^{-2b/3}$. 
After taking its derivative, we obtain the distribution of earthquakes in terms of energy:
\begin{equation}
 \label{G-R}
P(E) \sim E^{-\beta}, ~~ \text{with}~~\beta=2b/3  +1 = 5/3 = 1.67
  \end{equation}     

By considering that one earthquake with magnitude $M$ between 0 and 10 happens every 0.028 s, the distribution $P(E) \sim E^{-\beta}$ with $\beta=5/3$ corresponds to the red line of Fig.~\ref{Quakes}b, which reproduces well the worldwide distribution of earthquakes. 
In order to illustrate how important is the exponent in that real scenario, we analyse how different would be the predictions given by Eq.(1) considering  different exponent values and keeping constant the total number of events. 
The predicted number of earthquakes per year of a given magnitude as a function of $\beta$ appears in Fig.~\ref{Quakes}c.
 In the real of $\beta=5/3$ there is 1 catastrophic earthquake with magnitude between 8 and 9 (M8-9) per year (blue line). 
 However, in the case $\beta=3/2$, which corresponds to the mean field value of avalanche models \cite{Vespignani_1997} there would be 162.5 M8-9 earthquakes/year (almost one every second day).
 In the case of the critical situation of the OFC model of earthquakes ($\beta=1.27$) \cite{OFC_1992, Bonachela_2009}, there would be 138,000, roughly corresponding to a catastrophic M8-9 earthquake every 4 minutes.  
 These values have been calculated for a worldwide scenario, but the analysis is also valid at a more local level.
 Figure~\ref{Quakes}b shows that there is approximately one decade shift between the worldwide events and the Japan ones. 
 Thus by multiplying by 10 the average inter-event times just found earlier, we would have in average a catastrophic M8-9 earthquake every 10 years with $\beta=5/3$, every 22 days for $\beta=3/2$ and roughly every 40 minutes for $\beta=1.27$. 
 
 As the exponent $\beta$ decreases the differences in number between large events and small events decrease, vanishing at the value $\beta=1$, which corresponds to a b-value equal zero. In this situation the number of earthquakes as a function of the magnitude is no longer an exponential decay (as the Gutenberg-Richter law), but a constant (e.g., M8-9 earthquakes are as frequent as M2-3 ones). For $\beta$ values smaller than 1 the corresponding b-value is negative, which indicates that large magnitude earthquakes are more frequent that small magnitude ones.

The relations between frequent small earthquakes and very rare catastrophic ones are essential into understanding the nature of catastrophic earthquakes. As just shown, considering a different exponent value than $\beta=5/3$ may result in a scenario really far from the dynamics of real earthquakes. Unfortunately, many different experimental and theoretical models of earthquakes have not paid attention to this issue and it is possible to find reports with Gutenberg-Richter-like relations even with $\beta<1$.
 
How different is the physics when very large events are rare (as in the real earthquake scenario), or when they are quite abundant is a relevant question to focus on. In the first case the disordered interface would probably be the energy reservoir and after a very large quake the system may take a relative long time to recover before being ready for a new big one. In the second case, with large events happening frequently, the large amount of energy being continuously released at the interface may came from a large reservoir, as is the elasticity of the bulk in fracture experiments \cite{Bares2018}.
 

\bibliography{paperbib}